\pdfoutput=1
\documentclass[english,aps,prb,reprint,showpacs,superscriptaddress]{revtex4-1}
\usepackage{graphicx}
\usepackage{amsmath}
\usepackage{amssymb}
\usepackage{amsfonts}
\usepackage{dcolumn}
\usepackage{latexsym}
\usepackage{color}
\usepackage{latexsym}
\usepackage{subfigure}
\usepackage{float}
\usepackage{epsfig}
\usepackage{natbib}
\usepackage{bm}
\usepackage{amsthm}
\usepackage{eucal}
\usepackage{mathrsfs}
\usepackage{url}
\usepackage{braket}
\usepackage[T1]{fontenc}
\usepackage[latin9]{inputenc}
\usepackage{units}
\usepackage{soul}

\usepackage{color} 

\usepackage{hyperref}
\hypersetup{
colorlinks=true,final=true,
        linkcolor=blue,
        citecolor=blue,
        filecolor=blue,
        urlcolor=blue,
}

\newcommand{\LAO}{LaAlO$_3$ }
\newcommand{\STO}{SrTiO$_3$ }
\newcommand{\noise}{$\langle\delta R_{sheet}^2\rangle/ \langle R_{sheet}\rangle^2$ }
\newcommand{\tbkt}{$T_{BKT}$ }
\newcommand{\laosto}{LaAlO$_3$/\STO}
\usepackage{graphicx}
\usepackage{dcolumn}
\usepackage{bm}

\usepackage{babel}

\begin{document}
\title{Correlated Non-Gaussian phase fluctuations in \laosto heterointerface}
\author{Gopi Nath Daptary}
\affiliation{Department of Physics, Indian Institute of Science, Bangalore 560012, India}
\author{Shelender Kumar}
\affiliation{Department of Physics, Indian Institute of Science, Bangalore 560012, India}
\author{Pramod Kumar}
\affiliation{National Physical Laboratory, New Delhi 110012, India}
\author{Anjana Dogra}
\affiliation{National Physical Laboratory, New Delhi 110012, India}
\author{N. Mohanta}
\affiliation{Center for Electronic Correlations and Magnetism, 
Theoretical Physics III, Institute of Physics,
University of Augsburg, 86135 Augsburg, Germany}
\author{A. Taraphder}
\affiliation{Department of Physics, Indian Institute of Technology Kharagpur, W.B. 721302 India}
\author{Aveek Bid}
\email{aveek.bid@physics.iisc.ernet.in}
\affiliation{Department of Physics, Indian Institute of Science, Bangalore-560012, India}

\begin{abstract}
We probe the existence of large correlated non-Gaussian phase fluctuations in the vicinity of the superconducting phase transition in the conducting layer residing at the interface of \laosto heterostructures. The non-Gaussian fluctuations appear between the Berezinskii-Kosterlitz-Thouless  transition temperature $T_{BKT}$ and the mean field transition temperature $T_C$. Subsequent theoretical analysis reveals that non-Gaussianity arises predominantly due to the percolative transition of a Josephson coupled network of superconductors. Our results confirm that the superconductivity in this system is confined to two-dimensions. Our study of the non-Gaussian resistance fluctuation spectrum provides a novel means to explore the BKT-transition in two-dimensional inhomogeneous superconductors.
\end{abstract}

\maketitle
\section{Introduction}
  In a two dimensional superfluid, the vortices induced by thermal fluctuations appear as bound pairs below a characteristic temperature - the Kosterlitz-Thouless-Berezinskii transition temperature ($T_{BKT}$)~\cite{berezinski1973jetp, KT}.  The thermally activated unbinding of these vortex pairs at temperatures above $T_{BKT}$ induces phase slippage  leading to the onset of resistance and the eventual destruction of superconductivity in these materials~\cite{PhysRevLett.42.1165, RevModPhys.59.1001}. Unfortunately, conventional techniques that probe for signatures of BKT transition such as the measurement of discontinuity in the superfluid density near the transition temperature~\cite{PhysRevLett.107.217003,PhysRevLett.110.037002} cannot be applied in the case of low-dimensional heterostructures where the charge carriers are buried at an interface.  A case in point is the two-dimensional electron gas residing at the interface of \laosto heterostructures~\cite{Reyren31082007, hwang2012emergent}. While transport studies reported in this system seem to indicate that the superconductivity is two dimensional in nature with a gate voltage dependent  $T_{BKT}$~\cite{Reyren31082007, caviglia2008electric, :/content/aip/journal/apl/105/19/10.1063/1.4901940, PhysRevB.79.184502, transient}, the temperature variation of the pairing gap shows a BCS behaviour with $2\Delta_0/k_{B}T_{gap} = 3.4$~\cite{Richter2013}. Thus, there is an urgent need to develop alternate techniques that can unambiguously determine the dimensionality of the superconducting phase in such novel systems. 
  
Resistance fluctuations in three dimensional superconductors at temperatures higher than the transition temperature $T_C$ are strictly Gaussian in nature~\cite{PhysRevLett.89.287001, PhysRevB.47.11420}. In a previous related work we have shown that in ultra-thin film superconductors undergoing BKT transition, the resistance fluctuations measured near the critical temperature contain strong non-Gaussian components (NGC)~\cite{PhysRevLett.111.197001} due to the presence of long range correlations among the fluctuating vortices. The Central Limit Theorem guarantees that for uncorrelated random processes (in this case resistance fluctuations), the fluctuation statistics are Gaussian.   As the correlation length in the system begins to diverge - for example near a critical phase transition - the resultant time-dependent fluctuation statistics develops a strong non-Gaussian component~\cite{PhysRevLett.100.180601, PhysRevB.90.115153, PhysRevLett.111.197001}. Such NGC, typically detected in a material through the  measurement of higher order statistics of its resistance fluctuations, were found to be completely absent in three dimensional superconductors~\cite{PhysRevLett.111.197001}.   

In this paper, we present detailed experimental and simulation studies of the resistance fluctuations in \laosto heterostructures~\cite{hwang2012emergent,  doi:10.1146/annurev-matsci-070813-113437,  transient} around the superconducting transition regime to  look into the possibility of interacting vortices in two dimensions. To probe the existence of the NGC in the resistance fluctuations in this system, we study the higher order spectra of resistance fluctuations  which we quantify through their second spectrum(described later in the text in detail)~\cite{restle1985non, seidler1996dynamical}. We find evidence of correlated fluctuations near $T_{BKT}$ and show that the NGC in the resistance fluctuations appear most likely because of the percolative nature of the superconducting transition. Our results confirm that the superconducting phase in this system is indeed confined to two-dimensions.

\section{Results}
\subsection{Sample preparation}
Our measurements were performed on samples with 10 unit cells of \LAO grown by Pulsed Laser Deposition on TiO$_{2}$ terminated (001) \STO single crystal substrates. The fluence was 0.50 $J cm^{-2}$ and the target-substrate distance was 4.8 cm. In order to achieve uniform TiO$_{2}$ termination the SrTiO$_{3}$ substrates were pre-treated with standard buffer HF solution~\cite{kawasaki1994atomic}. This was followed by treating the substrates  for an hour at 830$^\circ$C in an oxygen partial pressure of 7.4 $\times$ 10$^{-2}$ mbar to remove any moisture and organic contaminants from the surface and also to reconstruct the surface so that pure TiO$_2$ termination was realized. This was followed by the deposition of 10 unit cells of LaAlO$_{3}$ at 800$^\circ$C at an oxygen partial pressure of 1x 10$^{-4}$ mbar.  Growth with the precision of single unit cell was monitored by the oscillations count using in-situ RHEED gun. HRXRD measurements confirmed the epitaxial nature of the growth. The thickness of one unit cell from these measurements was obtained to be 3.75~\AA~\cite{0953-8984-27-12-125007,cancellieri2011electrostriction}. 

\begin{figure}[t]
\begin{center}
\includegraphics[width=0.48\textwidth]{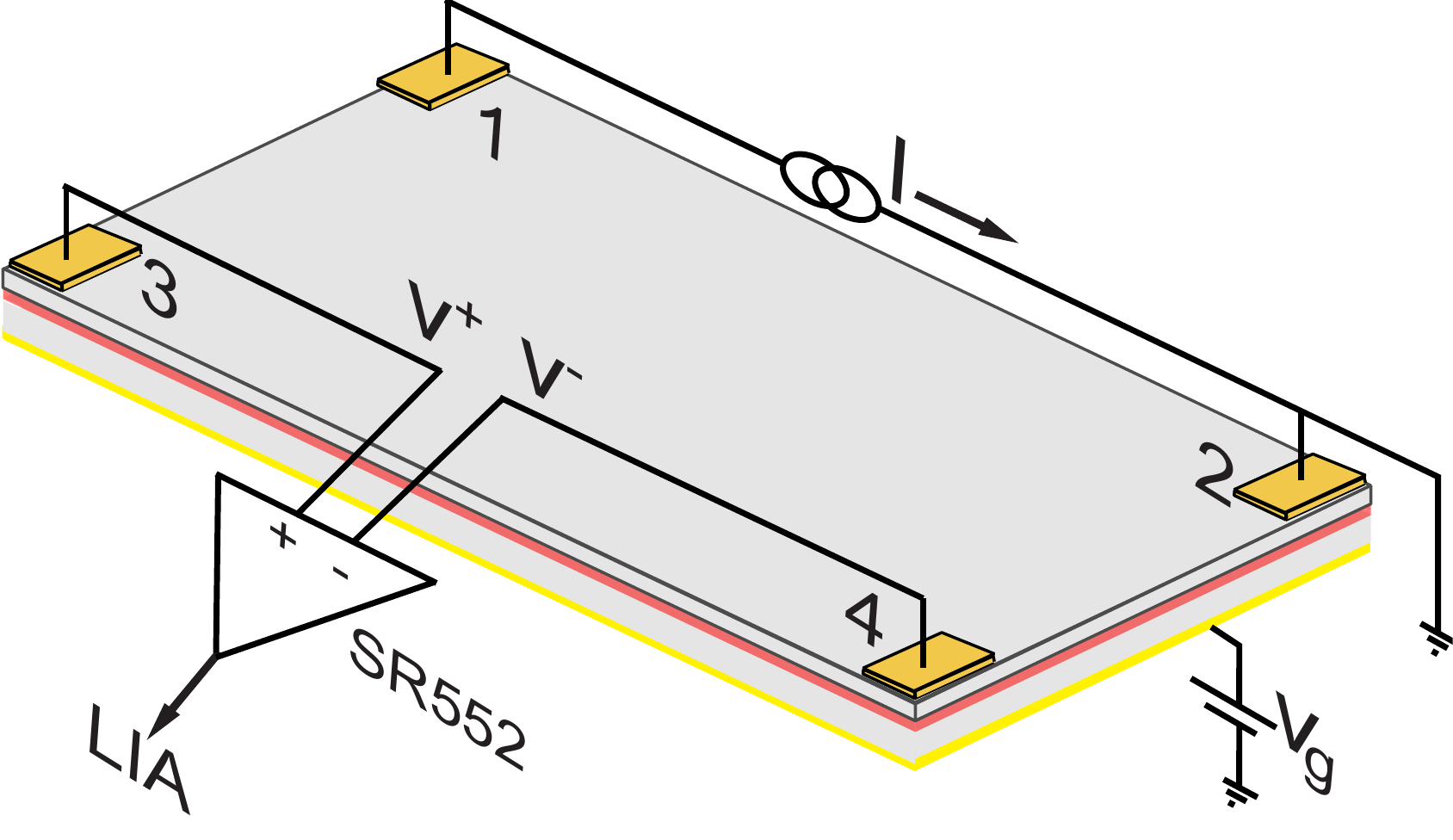}
\small{\caption{(color online) A schematic diagram of the device. The red shaded area represents the two-dimensional electron gas (2DEG) located at the interface of \LAO and SrTiO$_3$. The device is biased by a constant current between the contacts marked 1 and 2. The voltage difference between the contacts 3 and 4 is amplified by a low noise pre-amplifier (SR 552) and fed to a lock-in amplifier (SR 830) in differential mode. The carrier density is modulated by the back gate voltage $V_g$ applied to the contact at the bottom of the \LAO layer. The contact configuration for the measurement of resistance and resistance fluctuations were identical.
\label{fig:device}}}
\end{center}
\end{figure}

Electrical contacts were created on the heterostructures in van der Pauw geometry~\cite{rakhmilevitch2010phase, biscaras2010two, kalisky2012critical} by electron beam lithography followed by thermal evaporation of 5nm~Cr/100nm~Au contact pads.  The contact pads were connected to the chip carrier by ultra-sonic wire bonding which is known to breakdown the 10 u.c. of \LAO and give Ohmic contacts to the underlying electron gas ~\cite{caviglia2008electric,joshua2012universal,:/content/aip/journal/apl/105/19/10.1063/1.4901940,shalom2010tuning}. A typical device structure is shown schematically in figure~\ref{fig:device} along with the electrical connections. The charge carrier density level in the devices was controlled by a back gate voltage $V_g$ and was used to tune the system from a superconducting state to an insulating state. Sheet resistance, magnetoresistance (MR) and resistance fluctuation (noise) measurements were performed  over a wide range of $V_g$ and over a temperature range of 20 $mK$ to 400 $mK$ in magnetic fields up to 16 T in a cryogen-free dilution refrigerator.
 
\begin{figure}[tbh!]
\begin{center}
\includegraphics[width=0.48\textwidth]{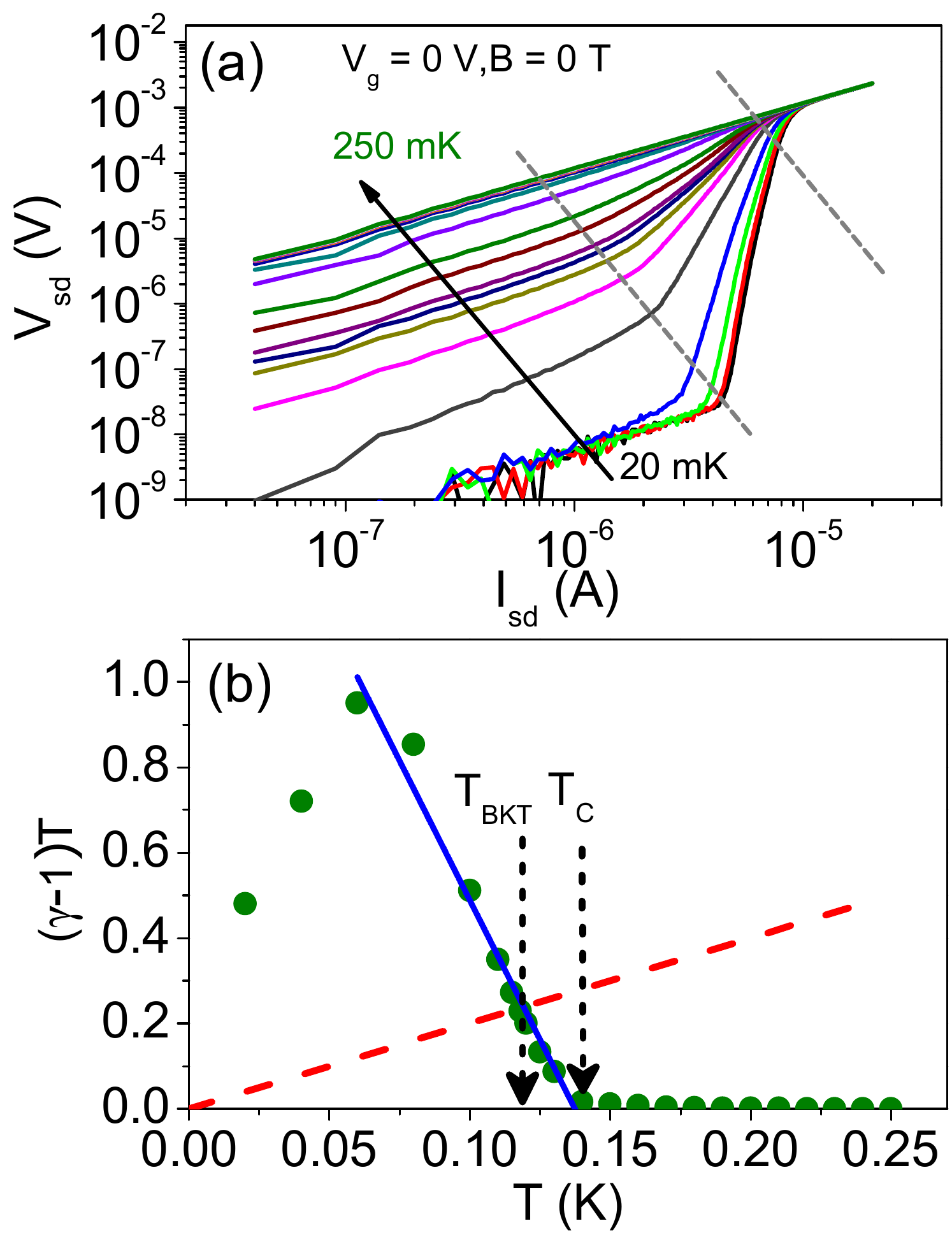}
\small{\caption{(color online)  (a) Plot of  the current \textit{vs} voltage characteristics of the sample at different values of temperatures ranging from 20 mK to 250 mK (from bottom: in steps of 20 mK from 20 mK to 100 mK then 110 mK, 115 mK, 118 mK, 120 mK, 125 mK and subsequently in steps of 10 mK from 130 mK to 250 mK). The exponent $\gamma$ is extracted by linear fit of the data for the current range indicated between the two grey dashed lines. (b) Plot of the scaled exponent $(\gamma-1)T$. The blue solid line is a fit to the data in the intermediate temperature range where the quantity ($\gamma-1)T$ is linear in temperature.  The intercept with the temperature axis determines the mean field transition temperature $T_C$. The red dotted line is the plot of $2T$ as a function of $T$ whose intercept with the measured data points determines $T_{BKT}$.
\label{fig:IV}}}
\end{center}
\end{figure}

\subsection{Extracting $T_{BKT}$ from electrical measurements} 
\begin{figure}[t]
\begin{center}
\includegraphics[width=0.48\textwidth]{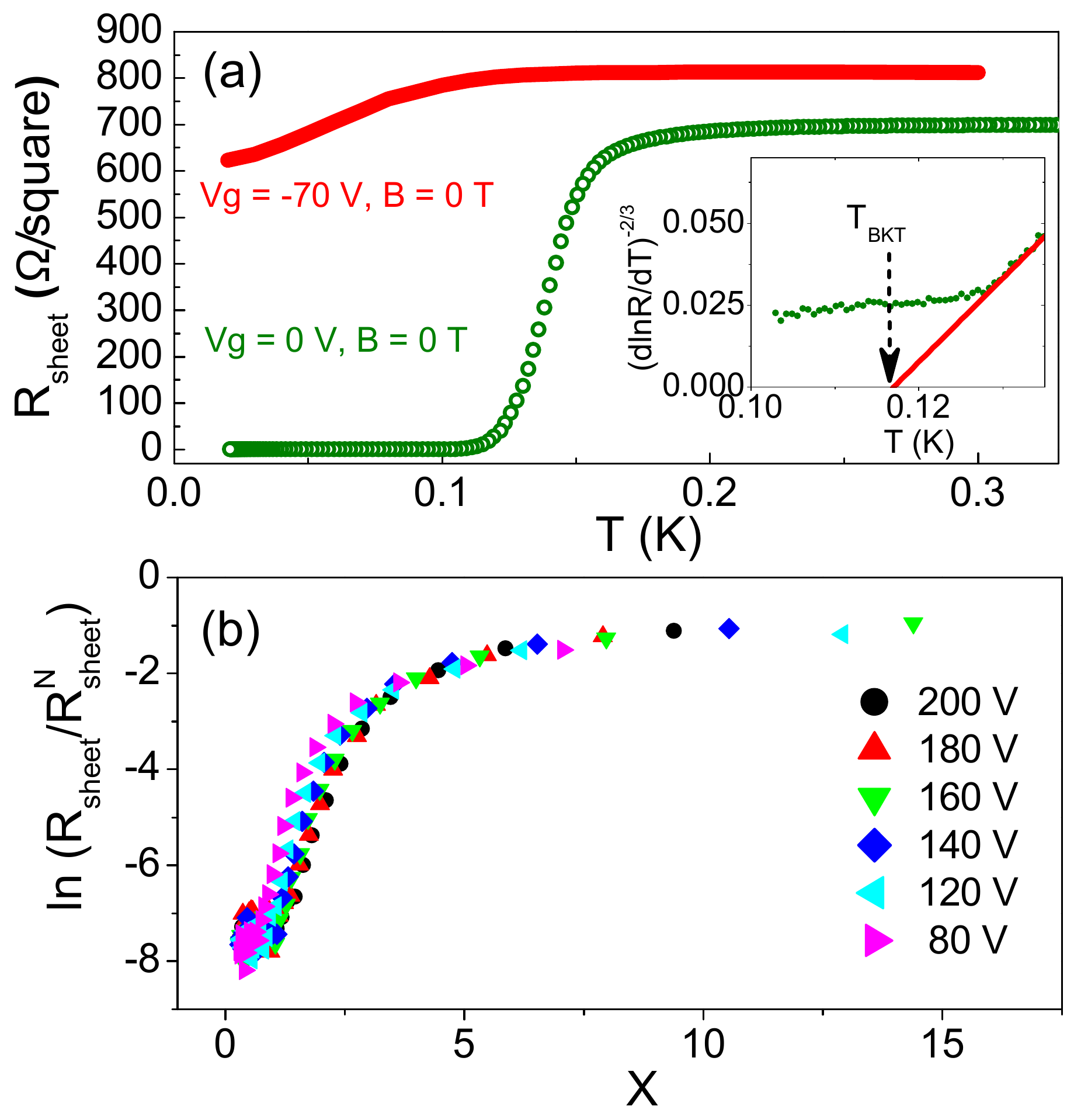}
\small{\caption{(color online)  (a) Sheet resistance as a function of temperature at two different values of $V_g$ - olive open circles:$V_g =0$~V, red solid line:$V_g=$-70~V. The inset shows a plot of $(dlnR/dT)^{-2/3}$ \textit{vs} temperature at $V_g =0$~V - the red line is the $BKT$ fit extrapolated to $(d ln R/dT)^{-2/3}$=0 to show explicitly the BKT temperature of 0.117~K. (b) Plot of  sheet resistance (normalized by the normal state resistance) as a function of the parameter $X = \frac{T}{T_{BKT}}\frac{(T_C-T_{BKT})}{(T_C-T)}$ obtained from measurements of resistance \textit{vs} temperature at six different values of $V_g$. The scaling of the data attests to the two-dimensional nature of the superconducting state.  The data were taken while passing a 100~nA ac current through the device.
\label{fig:RT}}}
\end{center}
\end{figure}

There are two main signatures of BKT transition from electrical measurements~\cite{PhysRevB.21.1806, PhysRevB.39.9708}.  The first  comes from measurements of non-linear current-voltage (IV) characteristics for $T<T_{BKT}$. According to the Ginzburg-Landau Coulomb gas description of a 2-D superconductor~\cite{PhysRevB.24.6758}, at temperatures below $T_{BKT}$ the application of a finite electrical current   should lead to the proliferation of free vortices inside the superconductor. The flux-flow resistance  generated by the flow of these free vortices is formally equivalent to a non-linear current-voltage relation: $V\sim I^{\gamma}$. The exponent $\gamma$ is temperature dependent and is described by the scaling relation~\cite{PhysRevB.28.2463, RevModPhys.59.1001} 
\begin{equation}
\gamma(T)-1 \sim \frac{1}{\epsilon(T)}\frac{(T_C-T)}{T}
\label{eqn:gamma} 
\end{equation}
Here $T_C$ is the Ginzburg-Landau (GL) mean field critical temperature. The term $\epsilon(T)$ is related to the polarization arising due to bound vortex pairs and can be approximated to 1 at sufficiently low temperatures~\cite{RevModPhys.59.1001}.  At the BKT transition temperature the non-linearity exponent $\gamma(T_{BKT})=3$~\cite{PhysRevB.28.2463, RevModPhys.59.1001,Reyren31082007,caviglia2008electric}. Note that this kind of power-law dependence of the IV characteristics is distinct from the exponential dependence of voltage on the current  arising from vortices motion due to flux depinning  seen in 3-dimensional superconductors. In figure~\ref{fig:IV}(a) we plot the IV curves measured over the temperature range 20~mK-250~mK at $V_g = 0$~V, the plots show strong non-linearity below 150~mK. In figure~\ref{fig:IV}(b) is plotted the function $(\gamma-1)T$ as a function of temperature. As expected from Eqn.~\ref{eqn:gamma} for a two-dimensional superconductor, the plot is approximately linear  down to about 60~mK. The deviation of the plot from linearity at very low temperatures has been observed previously in other systems~\cite{PhysRevB.28.2463} and has been attributed to possible flux pinning effects~\cite{PhysRevB.28.5075}.  The dotted line is a plot of 2$T$ against $T$. From the definition $\gamma(T_{BKT})=3$ it follows that the crossing of this line with the measured  $(\gamma-1)T$ data determines the BKT temperature.  Using this procedure  we identify $ T_{BKT} \sim$ 118~mK. An additional parameter extracted from such a fit is the mean field temperature $T_C$ which is determined by the intercept of the scaling plot on the temperature axis. For our device at $V_g=0~V$ the $T_C$ estimated by this method is $\sim$ 142~mK.

The second signature of BKT transition comes from the  temperature dependence of the resistance. Close to $T_{BKT}$, the resistance of a 2-dimensional superconductor depends on the temperature  as:
\begin{eqnarray}
R_{sheet}=R_0exp\Bigg(-\frac{b_R}{(T-T_{BKT})^{1/2}}\Bigg),
\label{eqn:resistance}
\end{eqnarray}
\noindent where $b_R$ is a measure of the vortex-antivortex interaction strength~\cite{PhysRevB.21.1806, PhysRevLett.55.2156, RevModPhys.59.1001}. This form is valid over a narrow range of temperature above $T_{BKT}$ where superconductivity is destroyed by phase fluctuations induced by  thermal unbinding of vortex-antivortex pairs.  Figure~\ref{fig:RT}(a) shows the sheet resistance of the device as a function of temperature at two characteristic gate voltages. To estimate $T_{BKT}$  we have fitted the data  using Eqn.~\ref{eqn:resistance} as shown in the inset of  figure~\ref{fig:RT}(a). The BKT transition temperature obtained from an extrapolation of this fit was $T_{BKT} \sim 117~mK$ for this particular device at $V_g=0~V$, in agreement with the value obtained from analysis of non-linear IV characteristics. Although this procedure is an useful method of extracting the $T_{BKT}$, it is at best an approximation and is as such not compelling enough to establish the 2-dimensional nature of superconductivity.  The biggest drawbacks of this method are - (i) it ignores finite size effects wherein the correlation length in the system is not allowed to grow beyond the sample size~\cite{PhysRevB.79.184502,PhysRevLett.28.1516, PhysRevB.30.322}, and (ii) it is strictly valid over a very narrow temperature window around $T_{BKT}/T$<1<$T_C/T$. To conclusively establish the 2-dimensional nature of superconducting state one needs to look at the scaling of the resistance with an appropriately normalized temperature scale~\cite{Halperin_JLTP1979}. In the absence of a magnetic field, the resistance of a 2-dimensional superconductor should be a universal function of its characteristic energy scales~\cite{Halperin_JLTP1979}. A particularly useful way of testing this prediction is by plotting the normalized resistance $R_{sheet}/R_{sheet}^N$ as a function of the scaling variable $X=T(T_C-T_{BKT})/[T_{BKT}(T_C-T)]$ using the values of $T_C$ and $T_{BKT}$ extracted from non-linear IV characteristics. The resulting plot should be independent of any additional sample parameters~\cite{RevModPhys.59.1001}. In figure~\ref{fig:RT}(b) we show the validity of this scaling relation by plotting the data obtained on our device over a range of values of $V_g$. 
The data from all the measurements collapse onto a single curve establishing that the superconductivity in this system is consistent with the GL Coulomb gas model for a 2-dimensional superconductor.  


\subsection{Resistance fluctuation measurements and analysis}
\begin{figure}[t]
\begin{center}
\includegraphics[width=0.48\textwidth]{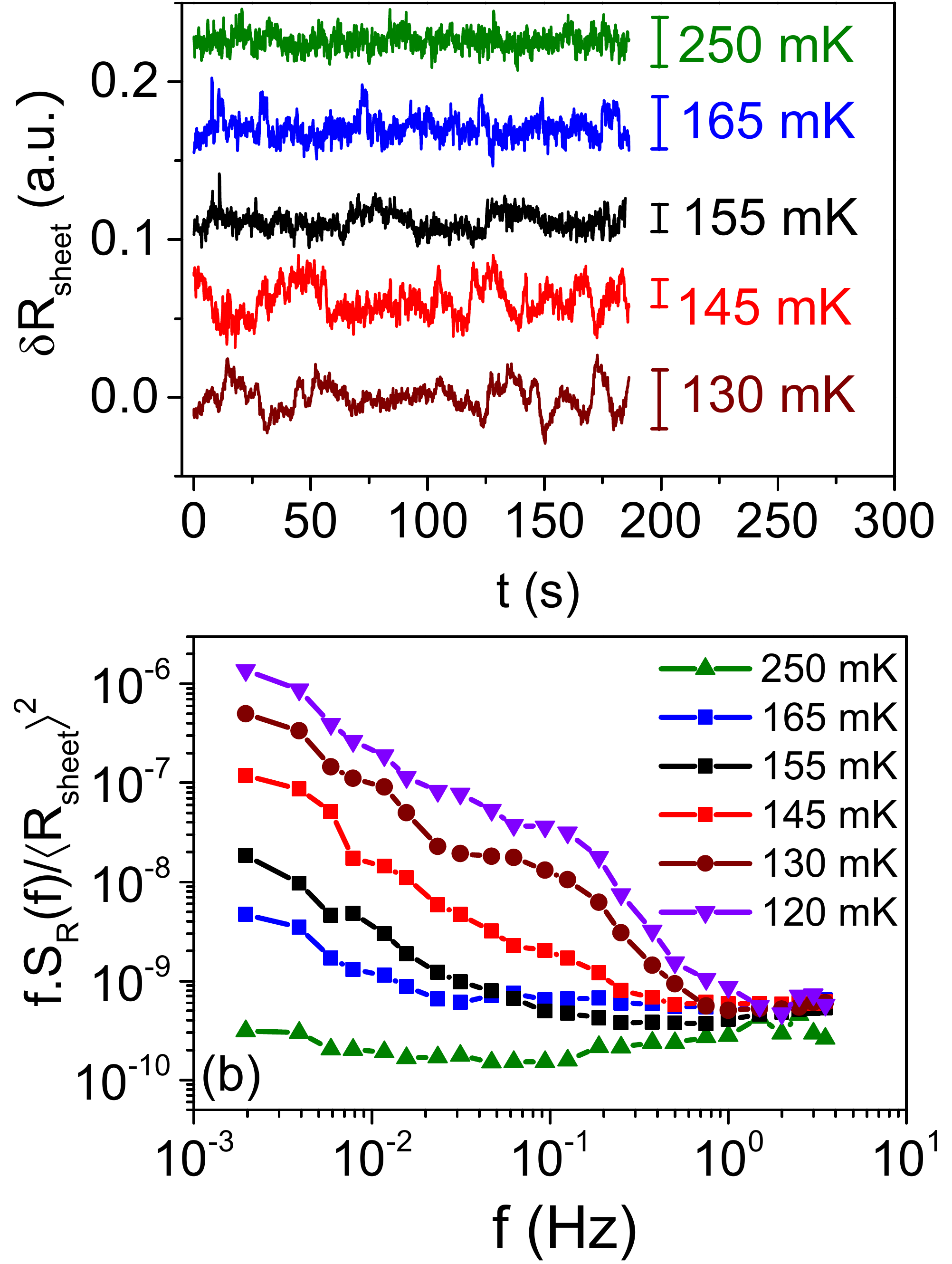}
\small{\caption{(color online) (a) Time-series of resistance fluctuations at a few representative temperatures. The vertical scale bar in each case represents 10 $m\Omega$. (b) Normalized power spectral density of resistance fluctuations $S_R (f)/\langle R_{sheet}\rangle^2$ plotted as a function of frequency calculated from the time-series shown in (a). Note that the data have been multiplied by $f$ to accentuate any deviation from $1/f$ nature of the spectrum.
\label{fig:time}}}
\end{center}
\end{figure}
To probe the statistics of fluctuations near the superconducting transition in this system we have studied in detail the low frequency resistance fluctuations at different gate voltages and magnetic field ranges using a  digital signal processing (DSP) based a.c technique. This technique allows simultaneous measurement of the background noise as well as the bias dependent noise from the sample ~\cite{scofield1987ac, Aveek_arxiv2004}. A low-noise pre-amplifier (SR-552) was used to couple the  sample to a lock-in-amplifier (LIA). The bias frequency of the LIA was chosen to lie in the eye of the Noise Figure (NF) of the pre-amplifier to minimize the contribution of the amplifier noise to the measured background noise.  The output of the LIA was digitized by a high speed 16 bit analog-to-digital conversion card  and stored in the computer. Typical time traces of resistance fluctuations as a function of time are shown in figure~\ref{fig:time}(a). The complete data set for each run, typically consisting of three million points, was decimated and digitally filtered to eliminate the 50~Hz line frequency component. The filtered time series of voltage fluctuations thus accumulated was used to calculate the power spectral density (PSD) of voltage fluctuations $S_{V}(f)$ over a specific frequency range. The lower frequency limit of this spectral  range ($~1 mHz$) was set by the stability of the temperature control which was better than $\pm 0.1~mK$. The upper cut-off frequency of the spectral range ($\sim 4 Hz$) was determined by the flatness of the response of the output low-pass filter of the Lock-in amplifier which had been set at 10~msec with a roll off of  24 dB/octave. The apparatus was calibrated using thermal noise measurements on standard resistors to measure spectral power densities down to $S_V \sim 10^{-20}~V^{2}Hz^{-1}$. The measured background noise was found to be bias independent, had a frequency independent spectrum and matched the value $4k_{B}TR_{sheet}$ expected for Johnson-Nyquist noise. The PSD of voltage fluctuations was converted to PSD of resistance fluctuations $S_R(f)$ using the relation $S_R(f) = S_V(f)/I^2$ where $I$ is the r.m.s. value of the constant current used to bias the device. At all temperatures measured, the dependence of $S_R(f)$ on the frequency $f$ was found to be of the form $S_R(f) \propto 1/f^\alpha$ - some representative plots are shown in figure~\ref{fig:time}(b). The value of the noise coefficient $\alpha$ ($=-\delta ln(S_R)/\delta ln(f)$) steadily increased from one at high temperatures to about three as $T_{BKT}$ was approached from above.  The high value of $\alpha$ very close to $T_{BKT}$ is indicative of the presence of percolative transport in this temperature regime~\cite{physC}, we address this point in detail later in this article.
\begin{figure}[t]
\begin{center}
\includegraphics[width=0.48\textwidth]{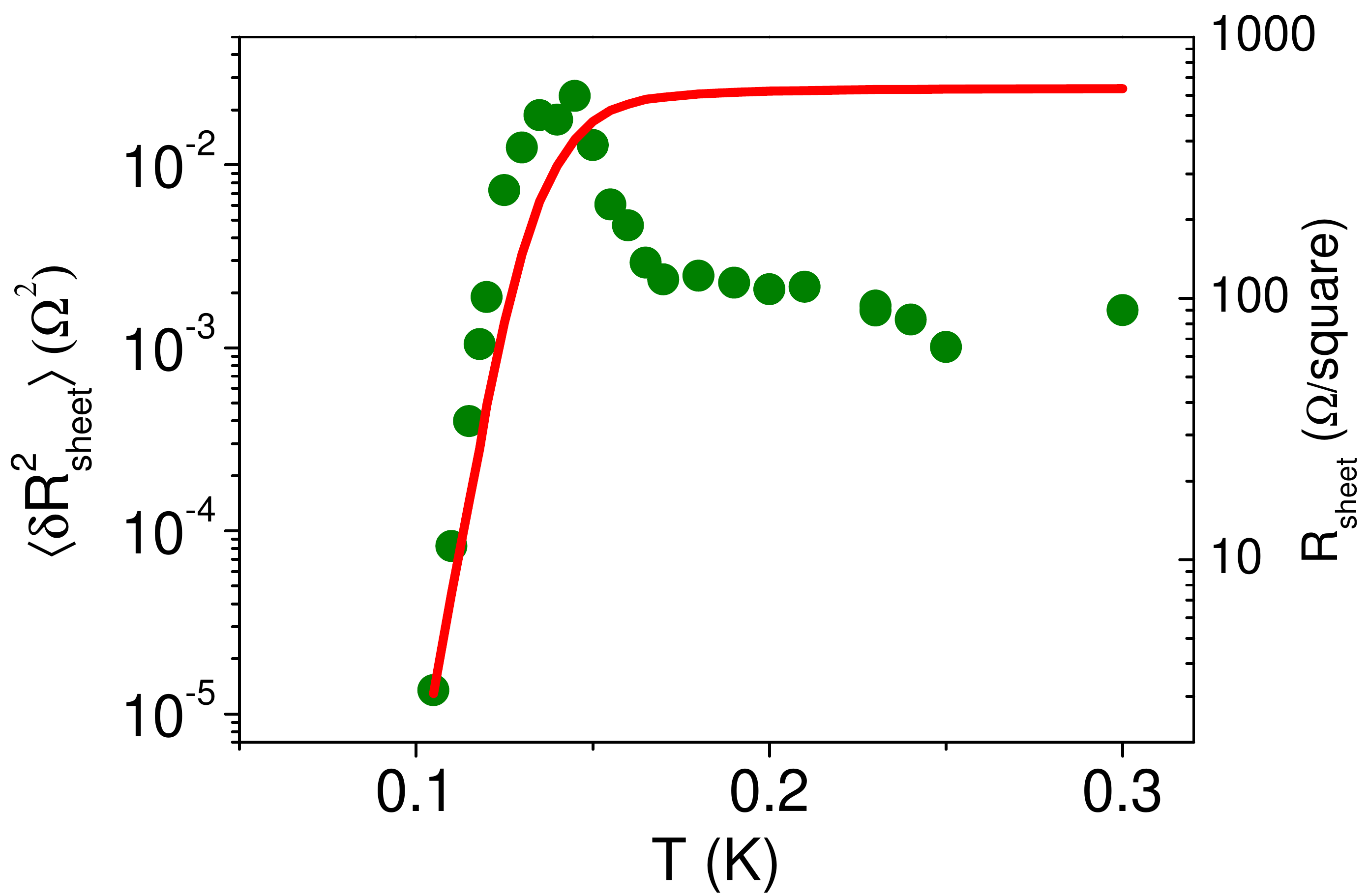}
\small{\caption{(color online) Plots of $\langle \delta R_{sheet}^2\rangle$ (left-axis - olive filled circles) and $R_{sheet}$ (right-axis - red line) as a function of temperature.
\label{fig:figure5}}}
\end{center}
\end{figure}
\begin{figure}
\begin{center}
\includegraphics[width=0.48\textwidth]{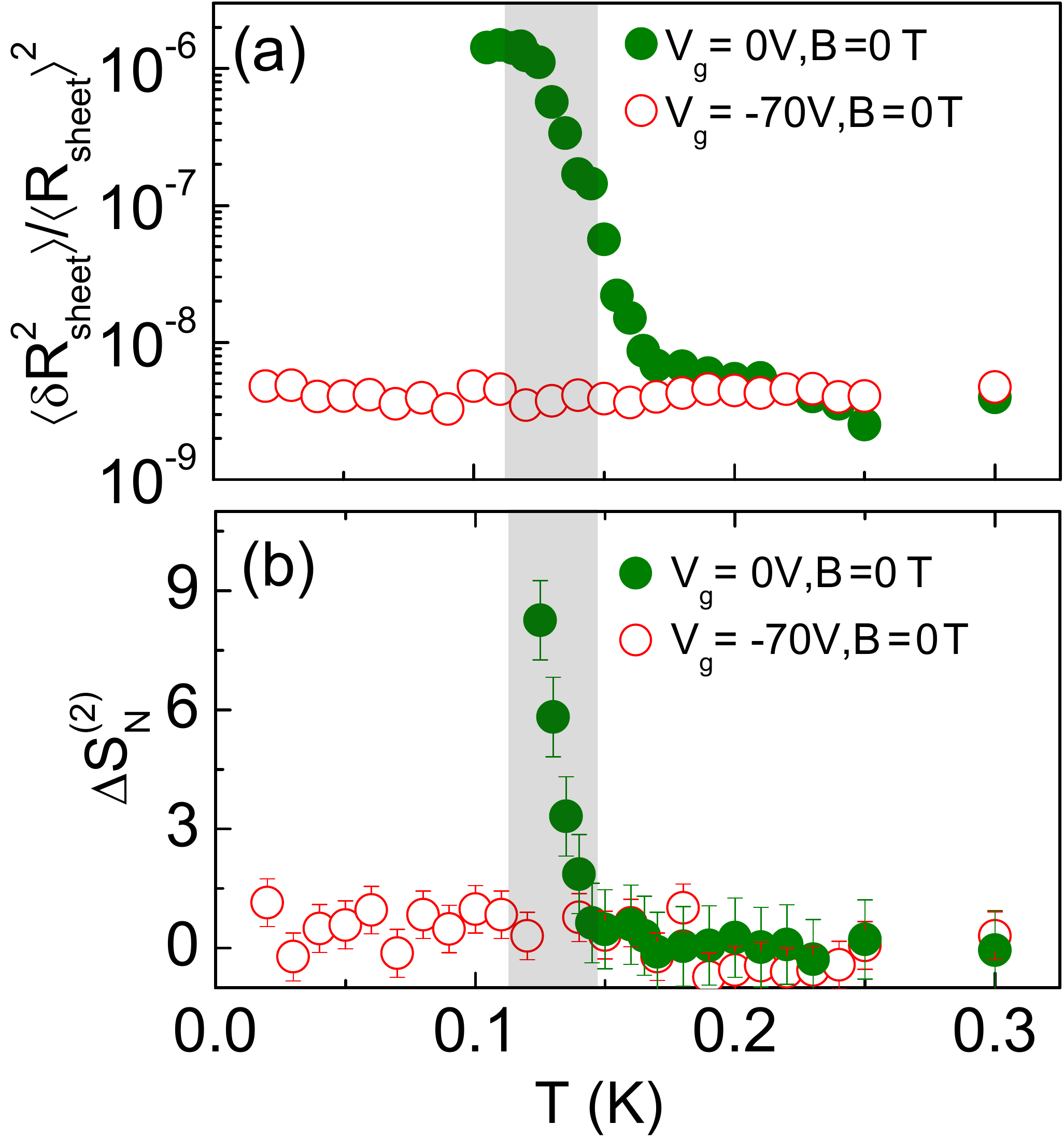}
\small{\caption{(color online) (a) Relative variance of resistance fluctuations as a function of temperature measured at $V_g=0$~V (olive filled circles) and $V_g=-70$~V (red open circles). Note the sharp increase in noise over the temperature range  $T_{C}>T>T_{BKT}$ (shown by the grey shaded area)  for $V_g$=0. The error in the data is smaller than the size of the symbols used. (b) Excess second spectrum $\Delta S_N^{(2)}$ as a function of temperature at different values of $V_g$ at zero magnetic field: $V_g=0$~V (olive filled circles), $V_g=-70$~V (red open circles). Error bars were calculated as standard deviations from measurements of $\Delta S_N^{(2)}$ over 50 time windows \label{fig:temp_noise}}}
	\end{center}
\end{figure}

The PSD of resistance fluctuations was subsequently integrated over the bandwidth of measurement to obtain the relative variance \noise of resistance fluctuations~\citep{Aveek_arxiv2004, scofield1987ac}:
\begin{eqnarray}
\frac{\langle\delta R_{sheet}^2\rangle}{\langle R_{sheet}\rangle^2}=\frac{1}{\langle R_{sheet}\rangle^2}\int{S_R(f)df}
\label{eqn:rv}
\end{eqnarray}
Plots of $\langle\delta R_{sheet}^2\rangle$ resistance as well as that of the sheet resistance are shown in Fig.~\ref{fig:figure5}.
Figure~\ref{fig:temp_noise}(a) shows a plot of  the relative variance of resistance fluctuations \noise measured as a function of temperature at $V_g = 0V$ and zero magnetic field. The magnitude of \noise increases by almost three orders of magnitude as the temperature approaches $T_{BKT}$. This large increase in noise near $T_{BKT}$ can be understood by considering the dynamics of the superconducting order parameter in the vicinity of the critical temperature. As the critical temperature is approached from above, fluctuations in the superconducting order parameter lead effectively to the formation of a dynamic network of superconducting and resistive regions in the sample. A major component of the divergent noise in this regime of phase space is understood to arise due to the fluctuations in the number/size of these superconducting domains~\cite{PhysRevB.38.2922, PhysRevB.61.1495}. This rapid increase of noise in the vicinity of superconducting transition has been reported before both in 2-D as well as in 3-D superconductors ~\cite{PhysRevLett.111.197001} and cannot help uniquely establish the dimensionality of the superconducting phase. 

We probed the presence of correlations in the system near the superconducting transition through the measurement of the second spectrum which is defined as the four point correlation function of the resistance fluctuations calculated over a chosen frequency octave ($f_l$, $f_h$)~\cite{restle1985non, seidler1996dynamical}. It is expressed mathematically as:
\begin{equation}
S_R^{f_1}(f_2)=\int_0^\infty \langle\delta R^2(t)\rangle\langle\delta R^2(t+\tau)\rangle cos(2\pi f_2\tau)d\tau
\label{eqn:2ndsp}
\end{equation}
where $f_1$ is the center frequency of the octave and $f_2$ is the spectral frequency. Operationally, the first step in calculating the second spectrum is to make repeated measurements of $S_R(f)$ over a frequency band and to form a `time series' of noise power. The power spectral of this time series then is a measure of  the fluctuations in the noise power within a frequency band of the original spectrum - this quantity is called the second spectrum. Thus, $S_R^{f_1}(f_2)$ physically represents the `spectral wandering' or fluctuations in the PSD with time. In the infinite time approximation, the estimated power at any frequency
should not vary with time and hence the second spectrum should be identically zero. However, due to the finite measurement time, each of the frequency components of the spectrum have a finite variance~\cite{PhysRevB.53.9753}.\\

 A convenient way of representing the second spectrum is through its normalized form $S_N^{(2)}$ defined as
 \begin{equation} S_N^{(2)}=\int_0^{f_h-f_l}S_R^{f_1}(f_2)df_2/[\int_{f_l}^{f_h}S_R(f)df]^2
 \label{eqn:norm2ndsp}
 \end{equation} 
 For Gaussian fluctuations, $S_N^{(2)}$=3 and any deviation from $3$ implies the presence of NGC in the fluctuation spectrum. 

We have calculated the second spectrum over the frequency octave 0.375~Hz - 0.750~Hz, where the sample noise is significantly higher than the background noise. This was in order to avoid corruption of the signal by the Gaussian background noise. The measured values of $S_N^{(2)}$ as a function of temperature at $V_g=0$~V are plotted in figure~\ref{fig:temp_noise}(b), note that we plot the `excess second spectrum' defined as $\Delta S_N^{(2)}=S_N^{(2)}-3$.  For $T\gg T_C$ $\Delta S_N^{(2)}  \sim 0$. As the temperature is decreased below $T_{C}$, $S_N^{(2)}$ starts increasing monotonically reaching a value of almost 12 near $T=T_{BKT}$. It is interesting to observe that the second spectrum reduces to the Gaussian value by 140~mK while the relative variance of resistance fluctuations \noise continues to evolve till at least 225~mK. This shows that the non-Gaussianity seen in the region dominated by vortex fluctuations has an origin distinct from the critical or percolative fluctuations that dominate the measured noise.   


\subsection{$V_g$ dependence of noise}     
\begin{figure}
\begin{center}
\includegraphics[width=0.48\textwidth]{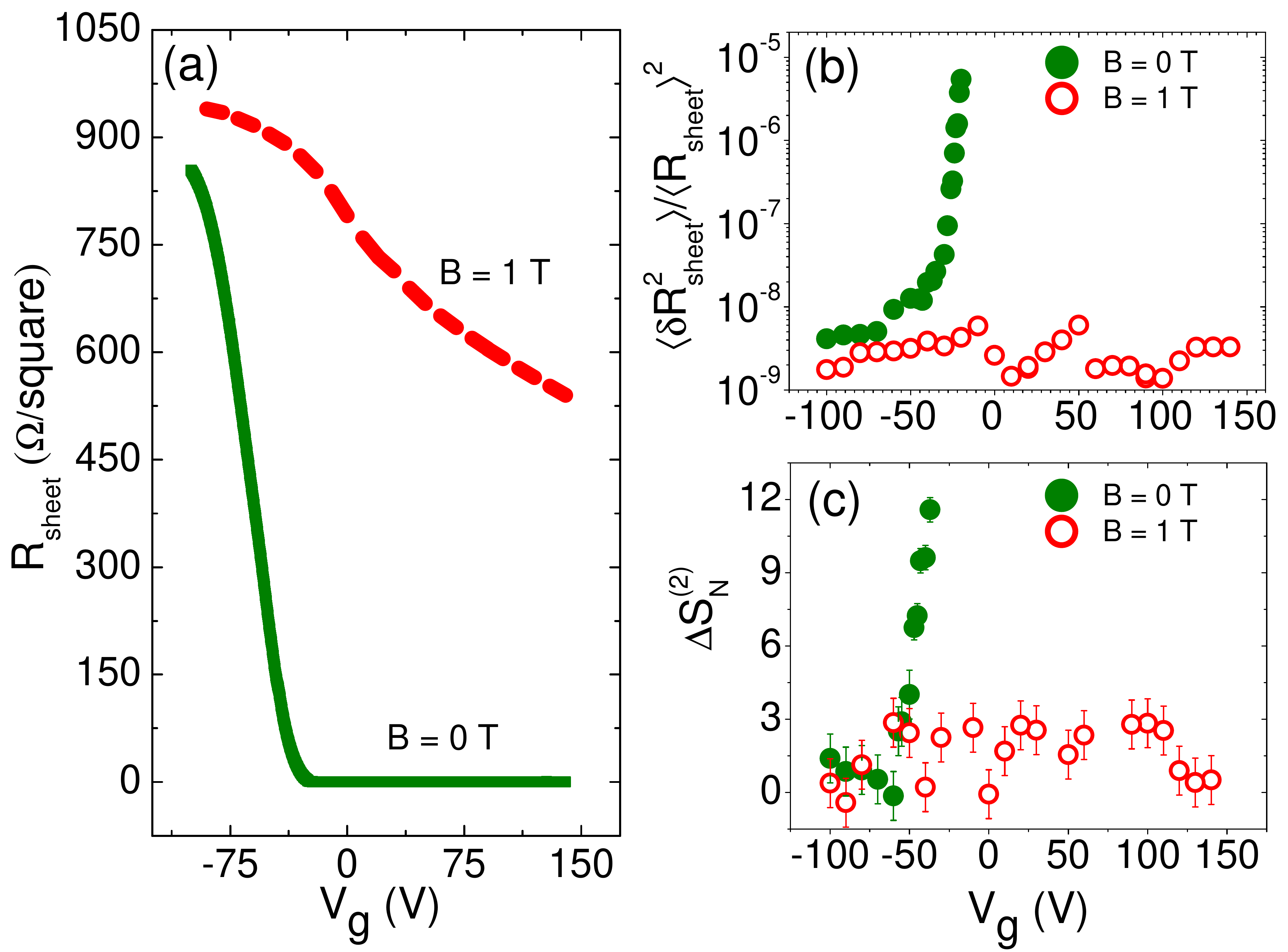}	\small{{\caption{(color online) (a) Plot of the sheet resistance as a function of $V_g$ at 0 T (olive solid line) and 1 T (red dashed line) magnetic fields . For $V_g > -20$~V the system is in a superconducting state at 20~mK and zero magnetic field. (b) Plot of the normalized noise \noise as a function of $V_g$ at 0~T (olive filled circles) and  1~T (red open circles) magnetic fields.  The error in the data is smaller than the size of the symbols used. (c) Normalized excess second spectrum $\Delta S_N^{(2)}$ as a function of gate voltage at 0~T (olive filled circles) and 1~T (red open circles) magnetic fields. Note that in zero magnetic field both the noise and $\Delta S_N^{(2)}$ increase rapidly as the superconducting transition is approached as a function of charge carrier density. Error bars were calculated as standard deviations from measurements of $\Delta S_N^{(2)}$ over 50 time windows. All the measurements were performed at 20 mK. \label{fig:gate_noise}}}}
	\end{center}
\end{figure}

We now turn to the effect of carrier density modulation on the resistance fluctuations in the superconducting state.  The superconducting state in this system can be tuned by modulating the carrier density using a back gate voltage $V_g$~\cite{Reyren31082007, caviglia2008electric, :/content/aip/journal/apl/105/19/10.1063/1.4901940, PhysRevB.79.184502, transient}. In figure~\ref{fig:gate_noise}(a) we plot the sheet resistance (green line) as a function of the gate voltage at 0~T magnetic field and 20~mK temperature. At this temperature the system showed superconductivity at all values of gate voltage larger than $V_g=-20$~V. For $V_g > -20$~V the PSD of voltage fluctuations $S_V(f)$ was below our detection limit of $S_V(f) \sim 10^{-20}~V^{2}Hz^{-1}$. As shown in figure~\ref{fig:gate_noise}(b), the measured noise \noise increased by more than three orders of magnitude as the system was driven from the normal state to the superconducting state by electronic doping. Concurrently, the value of $S_N^{(2)}$ increased monotonically from the Gaussian value of 3 near the normal state and reached a maximum near to the gate voltage at which the macroscopic resistance tends to zero - the data is plotted in figure~\ref{fig:gate_noise}(c). Application of a 1 T perpendicular magnetic field  suppressed the superconductivity at all gate voltages (figure~\ref{fig:gate_noise}~(a)), suppressed the measured noise (figure~\ref{fig:gate_noise}~(b)) and made the fluctuations Gaussian (figure~\ref{fig:gate_noise}~(c)). As a further control experiment we measured the noise at $V_g$ =~-70~V where the system is in a resistive state over the entire temperature range - the results are plotted in figure~\ref{fig:temp_noise}. As can be seen, the \noise is almost constant over the entire  temperature range and the normalized second spectrum $S_N^{(2)}$ shows a consistent Gaussian behavior. 

\section{Theory and Simulations}
In two-dimensional superconductors, the fluctuation in the conductivity arising from the proliferation of vortices at temperatures above $T_{BKT}$ is given by $\delta \sigma \propto \xi(T)^2$, where $\xi(T)$ is the coherence length at a temperature $T$. As pointed out in ref. ~\cite{benfatto} $b_R$ appears in the BKT correlation length $\xi$. The temperature-dependence of $\xi$ is given by the Halperin-Nelson formula~\cite{Halperin_JLTP1979}
\begin{equation}
\frac{\xi}{\xi_0}=\frac{2}{A}\sinh\frac{b_R}{\sqrt{T_r}}
\end{equation}
where $T_r=(T-T_{BKT})/T_{BKT}$ ($T \geqslant T_{BKT}$), $A$ and $b_R$ are parameters related to the vortex and determine the shape of the resistivity above $T_{BKT}$. The normalized resistance is given by
\begin{equation}
\frac{R}{R_N}=\frac{1}{1+(\Delta \sigma/\sigma_N)}=\frac{1}{1+(\xi/\xi_0)^2}
\label{Halperin-Nelson}
\end{equation}
where $R_N$ and $\sigma_N$ are, respectively, the resistance and conductivity in the normal state. 
However, the above temperature dependence is not sufficient to reproduce the extended tail usually observed in the sheet resistance curves near $T_{BKT}$ in inhomogenous two-dimensional superconductors. This tail, in fact, appears because of the percolative nature of the superconducting transition~\cite{PhysRevB.92.174531, Dubi2007, PhysRevB.75.184530}. At temperatures above $T_C$, there are superconducting islands in a non-superconducting background. These islands percolate via Josephson tunneling and grow in size as the temperature is reduced or as carriers are added to the system. There exists a characteristic temperature or doping at which macroscopic phase coherence is established throughout the entire sample and global superconductivity sets in. 

\begin{figure}[htb!]
\begin{center}
\includegraphics[width=0.4\textwidth]{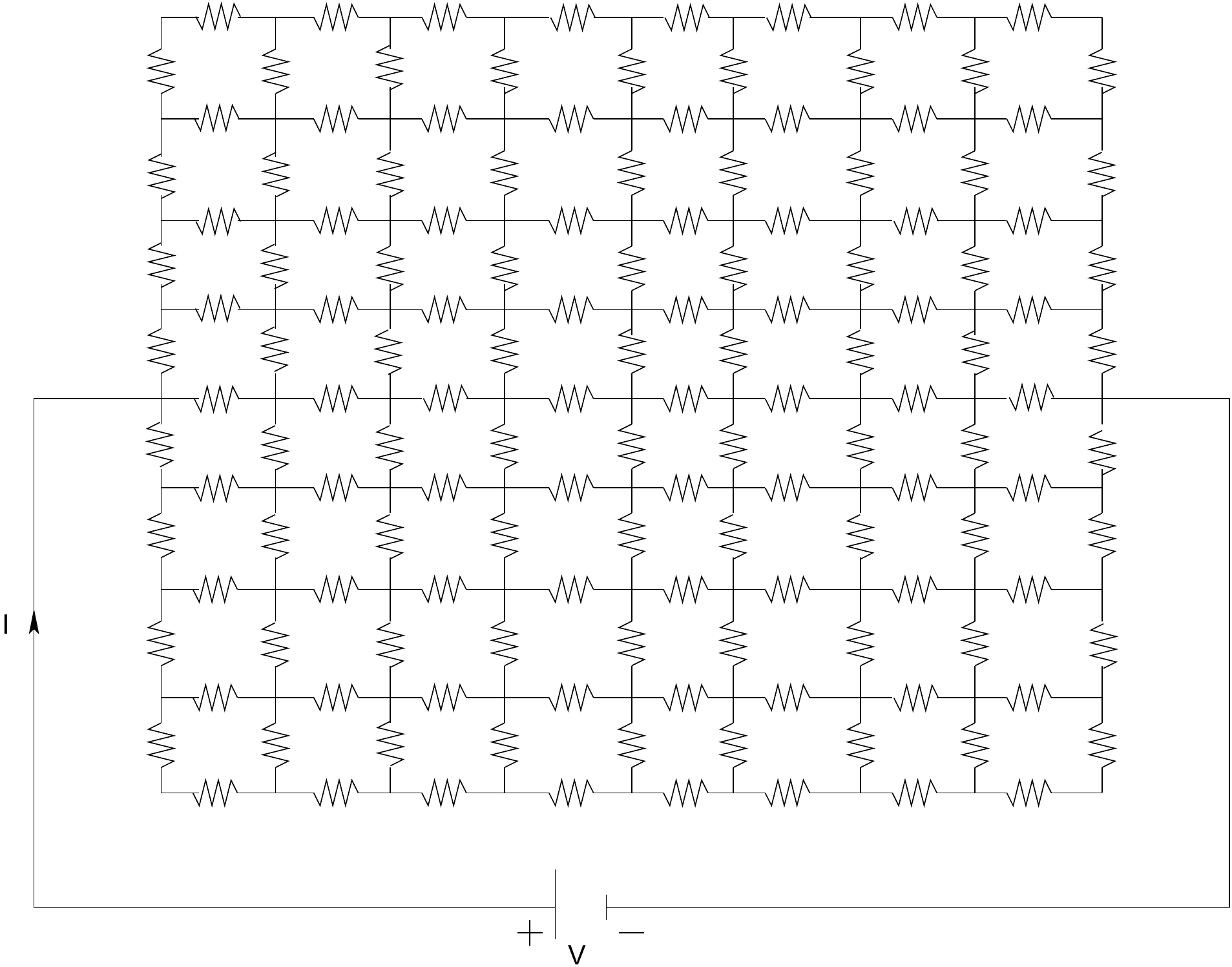}
\caption{(Color online) Schematic picture of a $9\times9$ resistor-network consisting of identical resistance at each junction. The macroscopic resistance $R$ is given by the current $I$ flowing through the outer wire driven by a potential $V$. A $100\times100$ network has been used in the calculation to model the percolation dynamics across the superconducting transition.}
\label{network}
\end{center}
\end{figure}

\begin{figure}[t]
\begin{center}
\includegraphics[width=0.5\textwidth]{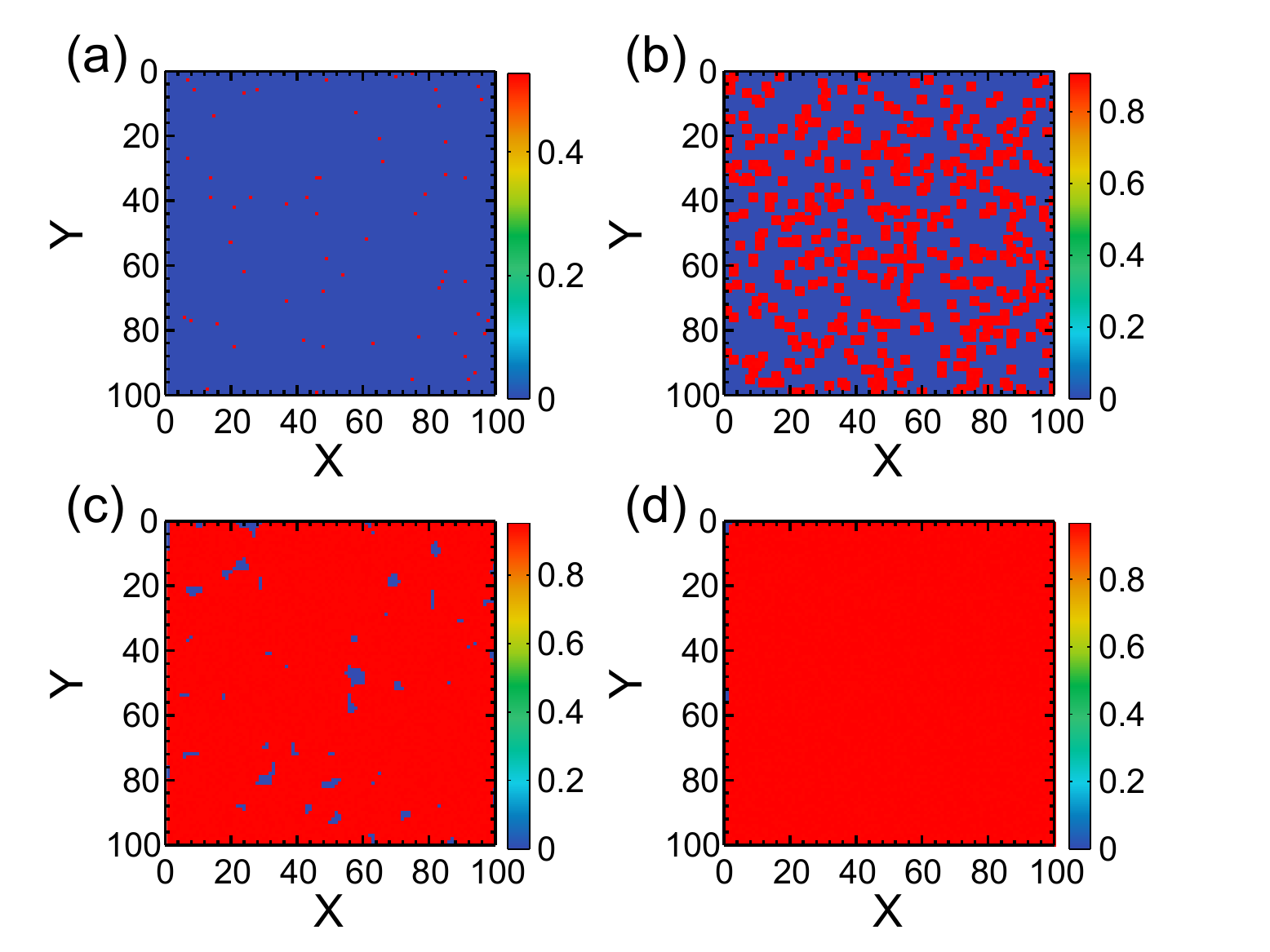}
\caption{(Color online) The spatial profiles of the normalized resistance in the resistor network of size $100\times100$ at temperatures (a) T=120 mK, (b) T=140 mK, (c) T=160 mK and (d) T=180 mK. The number and the diameter of the resistive patches both increase with temperature beyond $T_{BKT}$.}
\label{Res_maps}
\end{center}
\end{figure}

To gain insight into the origin of the NGC near the BKT transition, we perform a numerical simulation using a random resistor network (RRN) model.  We consider a two-dimensional square network of size $L\times L$ ($L$ being the number of junctions along one direction) with identical resistors present at all bonds of the network as described in figure~\ref{network}. The network is assumed to be connected by external conducting wires to a voltage source $V$ which causes a current $I$ to flow through the network, the macroscopic resistance is then given by $R=V/I$. However, for simplicity, we deal with the resistances at the junctions and the macroscopic resistance is taken as the average of all the junction resistances. To realize the percolation phenomenon occurring along with the superconducting transition, we include circular resistive patches with uniform superconducting background. As the temperature rises, the patches grow both in size and in number as depicted in figure~\ref{Res_maps}.
 
\begin{figure}[htb!]
\begin{center}
\includegraphics[width=0.48\textwidth]{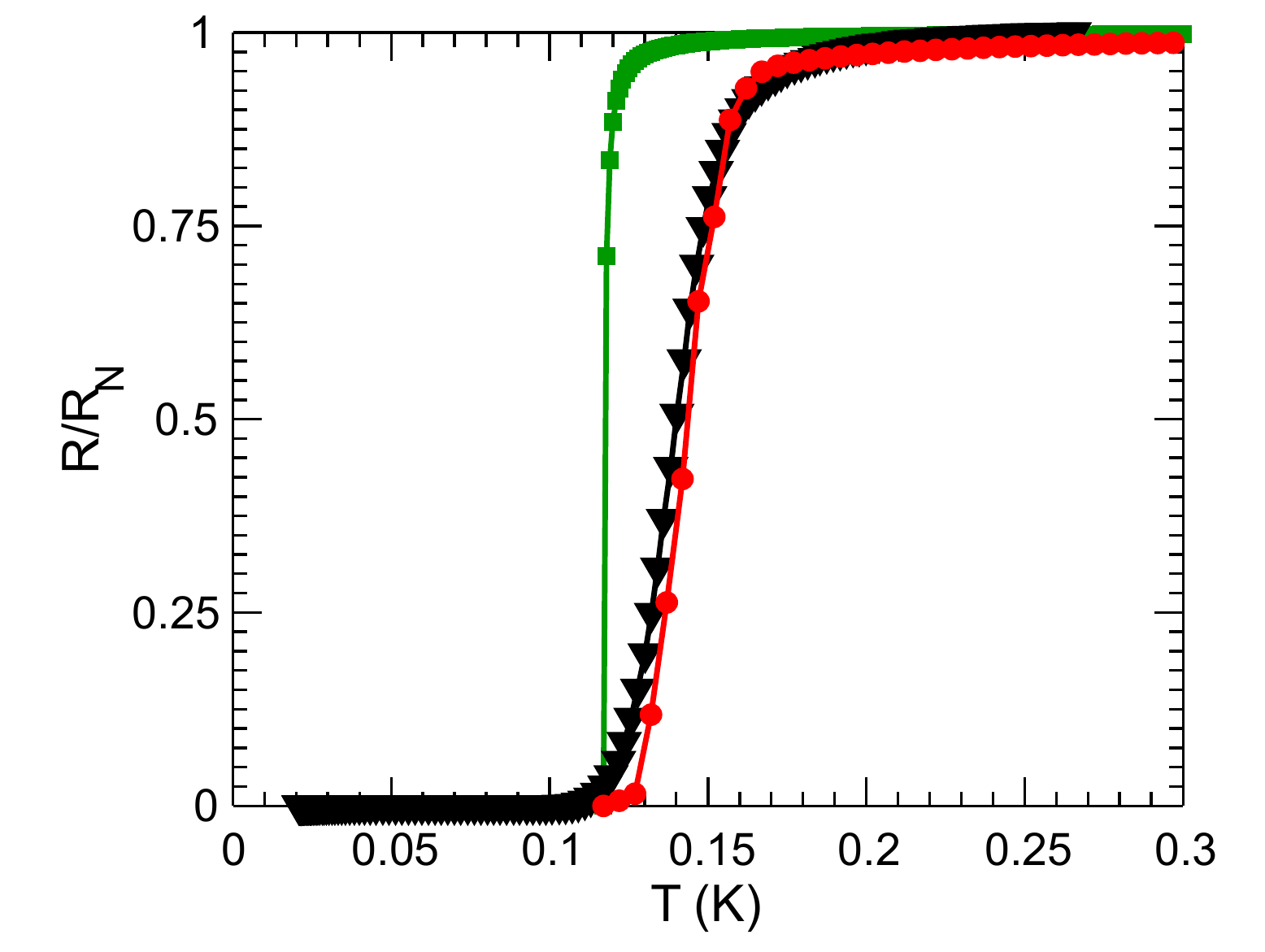}
\caption{(Color online) Plots of the  temperature dependence of the resistance calculated using using two different techniques - (i) Halperin-Nelson formula (olive squares) and (ii) average resistance of the $100\times100$ network extracted using the resistive patches to incorporate percolation (red filled circles). Parameters used are: $A=1.4$, $b_{R}=0.1$, $C_1=2.4$ and $C_2=10$. For comparison, the experimentally measured values of $R/R_N$ are also plotted (black filled triangles).}
\label{R_T}
\end{center}
\end{figure}

The resistance $r_j$ at a junction $j$ contained within a resistive patch at a temperature $T$ takes a value according to Eqn.~\ref{Halperin-Nelson}. The number and the diameter of the patches, at a temperature $T$, are given respectively by $N_{cluster}=C_1(T-T_{BKT})$ and $D_{cluster}=C_2T_r$. Here $C_1$ and $C_2$ are positive constants determined by matching the temperature variation of the average resistance with the experimental data as shown in figure~\ref{R_T}. For comparison we also plot on the same plot the experimentally determined value of $R/R_N$. 

Having defined the temperature dependence of the resistance at the individual junctions, we incorporate the fluctuations in the resistance. We start the simulation at temperature $T_{BKT}=117$ mK and reach $300$ mK and at each temperature, the network is set to evolve for a maximum time duration $t_{0}$. The resistance at the junction $j$ at a temperature $T$ and at an instant of time $t$ is given by
\begin{equation}
r_j (T,t)=r_j^0(T)+\Delta r_j(T,t)
\end{equation}
where $r_j^0(T)$ is the average (time-independent) value of  the $j^{th}$-resistance at temperature $T$ and $\Delta r_j(T,t)$ is the fluctuation around this value at a given instant of time $t$. To take into account the finite relaxation time $\tau$ of the resistance fluctuations, we start with the initial condition $\Delta r_j(T=T_{BKT},t=0)=0$ and update the resistance $r_j (T,t)$ continuously at each interval of time $\tau$. The set of fluctuations in resistances $\{ \Delta r_j(T,t)\}$ is obtained from a normal distribution with a standard deviation $0.001$ (dimensionless because the resistance $r_j$ is in the normalized form - see Eqn.\ref{Halperin-Nelson}) and having mean at zero. The relaxation time $\tau$ is also chosen randomly from a set $\{\tau_n\}$ which governs the statistics of the fluctuations in the resistance. At a given temperature, we use $\{\tau_n\}=x\{\tau_n\}_{NGC}+(1-x)\{\tau_n\}_{GC}$, where GC stands for Gaussian component and NGC for non-Gaussian component. The parameter $x$ controls the amount of NGC. Since the source of the NGC is the Josephson coupling of neighbouring superconducting islands, $x$ at a given temperature is taken to be proportional to the ratio of the superconducting region to the resistive region at that temperature. The exact values of the parameters determining the  distribution functions for $\{\tau_n\}_{NGC}$ and $\{\tau_n\}_{GC}$ are found by looking at the PSD, as discussed below. 

\begin{figure}[t]
\begin{center}
\includegraphics[width=0.48\textwidth]{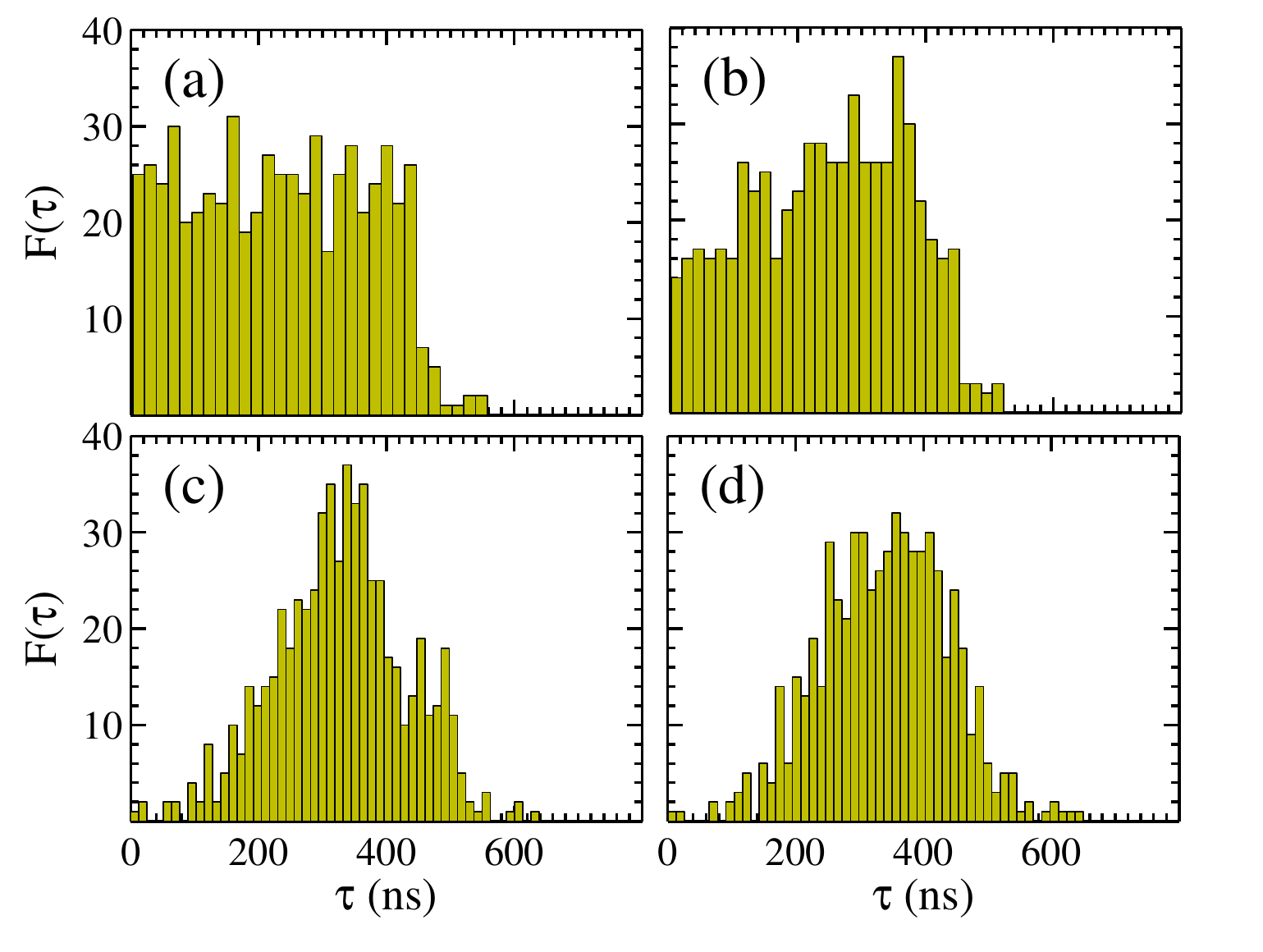}
\caption{(Color online) The distribution of the relaxation time $\tau$ at temperatures (a) $T=120$ mK, (b) $T=140$ mK, (c) $T=160$ mK and (d) $T=180$ mK. Gaussian distribution is evident at higher temperatures but deviation from Gaussianity appears primarily between $T_{BKT}=117$ mK and $T_c=150$ mK.}
\label{tau_dist}
\end{center}
\end{figure}

The PSD of the resistance fluctuation is given by
\begin{equation}
S_R(f)=\lim_{t_0 \to \infty} \Big(\frac{1}{2t_0}\Big) \Big(\int_{-t_0}^{t_0} \delta R(t)e^{-i2\pi ft}dt\Big)^2 
\end{equation}
and can be written in terms of the relaxation time $\tau$ as~\citep{Aveek_arxiv2004} 
\begin{equation}
S_R(f)=\int_0^{\infty} d\tau F(\tau) 2\tau/(1+(2\pi f \tau)^2) 
\label{psd_tau_eq}
\end{equation}
where $F(\tau)$ is the distribution function for $\tau$. The experimental data reveals that $S_R(f)\propto1/f^{\alpha}$, where $\alpha=1$ at higher temperatures and $\alpha=3$ at  $T \simeq T_{BKT}$. We input this information in the simulation through $F(\tau)$. For the NGC, we use the non-Gaussian distribution function $F(\tau)=(1/(2\sqrt{\pi}))\sqrt{\tau} e^{-\tau/\tau_0}$, typical for slow glassy dynamics~\cite{0034-4885-59-9-003}, with $\tau_0=600$ ns. For the GC, $F(\tau)$ is a Gaussian distribution with a mean $\tau_m=350$ ns and standard deviation $SD_{\tau}=100$ ns. The distribution of $\tau$ at different temperatures across the superconducting transition, revealing the dominance of the NGC within temperature range $T_{BKT} \leqslant T \leqslant T_{c}$, is shown in figure~\ref{tau_dist}. The distribution of the relaxation times, at a particular temperature, is extracted from the experimental data using the PSD, expressed in Eqn.~(\ref{psd_tau_eq}). While performing the time evolution, the relaxation time is taken randomly from the distribution.

\begin{figure}[tbh]
\begin{center}
\includegraphics[width=0.5\textwidth]{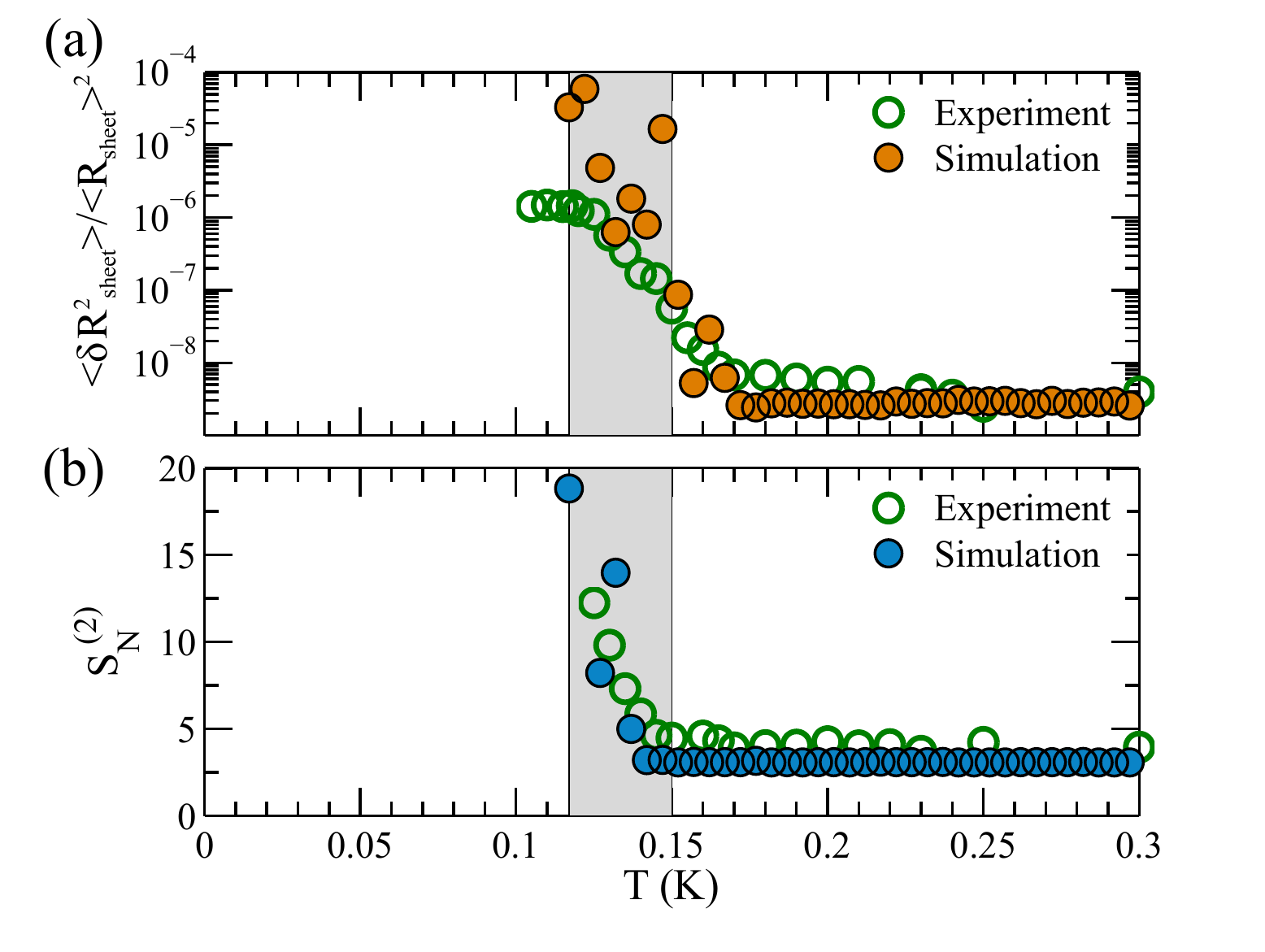}
\small{{\caption{(color online) Temperature variation of (a) the relative variance of the resistance fluctuations \noise (plotted in a semi-logarithmic scale) and (b) the normalized second spectrum $S_N^{(2)}$ obtained from the simulation of  the resistor-network model. For comparison the experimentally measured values of \noise and $S_N^{(2)}$ are also shown. 
\label{T_sigma}}}}
\end{center}
\end{figure}

The PSD of resistance fluctuations calculated above was used to estimate the second spectrum using Eqn.~\ref{eqn:2ndsp} and the normalized second spectrum was calculated by using Eqn.~\ref{eqn:norm2ndsp}. Since many random numbers are involved in the method, we perform several realizations of the random numbers and calculate the average values. To check the sensitivity of the final results to the initial conditions, we perform the calculation for several configurations of the initial resistance fluctuations with no qualitative difference in the final results.  In figure~\ref{T_sigma}, we plot the normalized noise \noise and the normalized second spectrum $S_N^{(2)}$, obtained from the simulation, as a function of temperature showing non-Gaussian fluctuations. The experimentally obtained values of \noise and $S_N^{(2)}$ are also plotted for comparison. The good match of the simulated data with the measured data attests to the fact that our simple model can capture the essential features of the resistance fluctuations near BKT transition in oxide heterostructures.       

 The simulation is composed of two parts - the temperature variation of the resistance, and the time evolution of the resistance at a particular temperature. The parameters $C_1$ (determines the number of the resistive clusters), $C_2$ (determines the radius of a cluster), the constants $A$ and $b_R$ in the Halperin-Nelson equation are tuned to match the temperature dependence of the resistance with the experimental data. At a given temperature, the time evolution of the resistance is set up by choosing the appropriate distribution functions using the parameters $\tau_0$ (governs the non-Gaussian distribution), $\tau_m$ (mean of the Gaussian distribution) and $SD_{\tau}$ (standard deviation of the Gaussian distribution).  

In a true BKT transition, vortex-fluctuations, above $T_{BKT}$, lead to the unbinding of paired vortices. Percolation adds extra fluctuations above $T_{BKT}$. Consequently, the $T_{BKT}$-$T_{C}$ range is increased and a tail appears in the temperature-versus-resistance plot. To compare the situations in BKT and non-BKT scenarios, we performed simulations using the same resistance network and relaxation time distributions considering BKT and BCS -type transitions and found that the large non-Gaussian fluctuation appears within the temperature range $T_{BKT}$-$T_{C}$ in case of BKT transition, and only in the vicinity of $T_C$ in case of BCS theory.


To conclude, we have probed the higher order statistics of resistance fluctuations around the transition temperature in the two dimensional superconducting state in \laosto interface. We find large non-Gaussian components in the fluctuation near \tbkt that signify strong correlations among interacting vortices. Our results confirm that the superconducting transition in this system is universal BKT-type in nature. Our theoretical simulation indicates that the large non-Gaussian resistance fluctuations are a manifestation of a percolative transition of a Josephson-coupled superconducting network.  Our analysis also suggests that the NGC in resistance fluctuation is a generic feature of two-dimensional inhomogeneous superconductors close to the transition temperature.

\begin{acknowledgments}
The authors thank R C Budhani for providing the samples. AB acknowledges funding from Nanomission, Department of Science \& Technology (DST) and the Indian Institute of Science.
\end{acknowledgments}


%


\end{document}